\documentclass[twocolumn,showpacs,amsmath,amssymb,pra,superscriptaddress]{revtex4}

\usepackage{graphicx}
\usepackage{amsfonts}
\usepackage{amscd}
\usepackage{amsmath}    
\usepackage{enumerate}

\newcommand{\bra}[1]{\mbox{$\left\langle #1 \right|$}}
\newcommand{\ket}[1]{\mbox{$\left| #1 \right\rangle$}}

\usepackage{amssymb}
\usepackage{amsthm}

\newtheorem{lemma}{Lemma}

\newtheorem{claim}{Claim}

\newtheorem*{observation}{Observation}
\newtheorem{definition}{Definition}

\begin{document}

\title{Practical issues in quantum-key-distribution post-processing}

\author{Chi-Hang Fred Fung}
\email{chffung@hku.hk}
\affiliation{Department of Physics and Center of Theoretical and Computational Physics, University of Hong Kong, Pokfulam Road, Hong Kong}%
\author{Xiongfeng Ma}
\email{xfma@iqc.ca}
\affiliation{%
Institute for Quantum Computing and Department of Physics and Astronomy, \\
University of Waterloo, 200 University Ave W., Waterloo, ON, Canada N2L 3G1 \\
}%
\author{H. F. Chau}
\email{hfchau@hkusua.hku.hk}
\affiliation{Department of Physics and Center of Theoretical and Computational Physics, University of Hong Kong, Pokfulam Road, Hong Kong}%

\begin{abstract}
Quantum key distribution (QKD) is 
a secure
key generation method between two distant parties by wisely exploiting
properties of quantum mechanics.
In QKD,
experimental measurement outcomes on quantum states
are transformed by the two parties to a secret key.
This transformation is composed of many logical steps (as guided by security proofs), which together will ultimately determine the length of the final secret key and its security. 
We detail the procedure for performing such classical post-processing taking into account practical concerns (including the finite-size effect and authentication and encryption for classical communications).
This procedure is directly applicable to realistic QKD experiments, and thus serves as a recipe that specifies what post-processing operations are needed and what the security level is for certain lengths of the keys.
Our result is applicable to the BB84 protocol with a single or entangled
photon source.

\end{abstract}


\maketitle

\section{Introduction}
In theory, a few quantum key distribution (QKD) protocols, such as BB84~\cite{BB_84}, BBM92~\cite{BBM_92}, B92~\cite{Bennett_92} and six-state~\cite{sixstate_98,Gisin:six:99}, have been proven to be unconditionally secure in the last decade \cite{Mayers1996,Mayers_01,LoChauQKD_99,ShorPreskill_00,Lo:six:01,Tamaki:B92:03,Renner_Thesis_05}. Security of other protocols, such as the Ekert91 protocol~\cite{Ekert_91} and the device-independent QKD protocol \cite{Acin:DeviceIn:07}, have also been studied. For a review of QKD, one can refer to~\cite{GRTZ_02,LoLut_review_07,RevQKD_Lut_08}.

QKD schemes can be classified into two types: prepare-and-measure scheme and entanglement-based scheme. In the former, one party, Alice, prepares the quantum signals (say, using a laser source) according to her basis and bit values and sends them through a quantum channel to the other party, Bob, who measures them upon reception.
In the latter type, an entanglement source emits pairs of entangled signals, which are then measured in certain bases chosen by Alice and Bob separately.
There is an important difference in terms of security between the emitted signals with practical sources in the two cases. In the prepare-and-measure case, the signals emitted by Alice (say, a weak coherent-state source) is basis-dependent, meaning that the coherent-state signal corresponding to one basis is quantum mechanically different from that of the other basis. An eavesdropper, Eve, can certainly leverage this information to her advantage. New techniques such as the decoy-state method~\cite{Hwang_03,Decoy_05,Practical_05,Wang_05,Wang2_05,HEHN_05,ZQMKQ_06,ZQMKQ60km_06}, strong-reference-pulse QKD~\cite{Tamaki:B92:04,Koashi_StrRef_04,TLMB_Strong06,Tamaki:Strong:08}, and DPS~\cite{IWY_DPS02,IWY_DPS03,TDLFY_10GHz_06,Takesue2005} have recently been invented to 
allow the use of coherent-state sources securely.
On the other hand, entanglement-based QKD involves signals that are basis-independent. The security of basis-independent QKD (including entanglement-based QKD with a PDC source and prepare-and-measure QKD with a single-photon source) has been proven in Ref.~\cite{KoashiPreskill_03} and the performance of entanglement-based QKD with a PDC source has been analyzed recently \cite{EntanglementPDC_07}.

As security analysis of QKD has become mature,
it now comes to the stage to consider all the underlying assumptions and to apply the theoretical results to practical QKD experiments.
Although standard security proofs (such as Ref.~\cite{ShorPreskill_00}) imply a procedure for distilling a final secret key from measurement outcomes, such a procedure cannot be directly carried out in an {\em actual} QKD experiment 
because
many of the security proofs focus on the case that the key is arbitrarily long. 
Although, in theory, there is not a fundamental limit on the length, it is constrained by the computational power in practice.
Therefore, it is imperative to quantify the finite-size effect and to provide a precise post-processing recipe that one can follow for distilling final secret keys with quantified security in real QKD experiments.
This is the motivation of this paper.
We remark that it is not that the security proofs are incorrect, but carrying them out in practice requires more additional consideration on the relation between the actual steps taken and the final security parameter.

Ultimately, QKD system designers would like to know the classical computation and communications needed to transform the measurement results of a QKD experiment to a final key.
Furthermore, it is important to know the trade-off between the final key length and the security parameter, since
this allows one to estimate the number of initial quantum signals to be sent in order to achieve a certain final key length and security.
We provide the solution to this in the current paper.

It is important to note that the post-processing procedure contains many elements including authentication, error correction and verification, phase error rate estimation, and privacy amplification.
Integrating all these elements with a security proof is nontrivial and the resultant post-processing procedure is the main contribution of our paper.
Our paper uses the latest security proof techniques to perform post-processing analysis, along a similar line to an early work by L\"utkenhaus~\cite{Lutkenhaus:practical:99}.
We emphasize that the main focus of our work is the overall procedure for post-processing, rather than analyzing a security proof in the finite-key situation.
We note that, recently, lots of efforts have been spent on the finite-key effect in QKD post-processing, such as Refs.~\cite{Hayashi:Finite:06,Scarani:Finite:08,Scarani:Finite:2008b,Cai_finite_2008}.

The finite-key-length analysis is important not only from a theoretical point view, but also for experiments. For example, the efficient BB84 \cite{EffBB84_05} is proposed to increase the key generation rate. In order to select an optimal bias between the two bases, $X$ and $Z$, Alice and Bob need to consider statistical fluctuations. We will address this issue in this paper.
We remark that the proposed post-processing scheme ties up a few existing results with some modifications. 
Key features of our work are as follows:
\begin{enumerate}

\item
a strict bound for the phase error estimation is derived;

\item
an authentication scheme is applied for the error verification;

\item
the efficiency of the privacy amplification is investigated;

\item
parameter optimization is studied.
\end{enumerate}
This paper is an expansion of our shorter paper~\cite{Finite:Short:09} which summarizes the essential components of our data post-processing procedure.
All the technical details related to the procedure are presented here.

The paper is organized as follows. 
Sec.~\ref{Sc:Assum} introduces the assumptions used in the paper. 
Sec.~\ref{sec-method} discusses the security aspect of our procedure.
Sec.~\ref{Sc:Outline} outlines the post-processing procedures. 
Sec.~\ref{sec-preliminary} introduces some preliminary tools to be used in later sections.
Secs.~\ref{sec-basis-sift}-\ref{sec-phase-error-correction} discuss the details of the various post-processing steps.
In Sec.~\ref{Sc:Optim}, we investigate the parameter optimization problem for this post-processing procedure.
In Sec.~\ref{Sc:Simul}, we simulate an experiment setup as an example.
We conclude in Sec.~\ref{sec-conclusion}.

\section{Assumptions} \label{Sc:Assum}
Here, we examine the underlying assumptions used in the post-processing scheme we propose in this paper. We emphasize that in order to apply the scheme to a QKD system, one needs to compare these assumptions with the real setup.
The assumptions used in the paper are listed as follows:

\begin{enumerate}
\item
Alice and Bob perform the BB84 protocol with a perfect single photon source (or basis-independent photon source \cite{EntanglementPDC_07});

\item
The detection system is compatible with the squashing model~\cite{TT_Thres_08,BML_Squash_08} (see also Ref.~\cite{Koashi_NewModel_06}). 
For example, the efficiency mismatch is not considered in this paper;

\item
Alice and Bob use perfect random number generators and perfect key management. They share a certain amount of secure key prior to running their QKD system.

\end{enumerate}

\section{Security aspect\label{sec-method}}

Our data post-processing procedure is derived from entanglement distillation protocol (EDP)-based security proofs~\cite{LoChauQKD_99,ShorPreskill_00,Lo_decouple_2003} (also see~\cite{Koashi_Uncer_06}) and thus our procedure is secure against the most general attacks allowed by the laws of quantum mechanics. The original idea~\cite{LoChauQKD_99} casts QKD as distilling Einstein-Podolsky-Rosen (EPR) pairs between Alice and Bob, involving correcting general quantum errors.
And the ability to correct general quantum errors is equivalent to the ability to correct bit and phase errors~\cite{Gottesman1996,Gottesman_Thesis_97}.
Later, Shor and Preskill~\cite{ShorPreskill_00} show that correcting bit errors and phase errors in the EDP picture correspond to bit error correction and privacy amplification in distilling a secret key. Thus, proving the security of QKD can be cast as showing that both bit and phase errors are corrected in the EDP picture. In this way, provided that a quantum error correction code 
exists for the specific bit and phase error rates in the EDP picture, the security of the corresponding QKD protocol is proved.
However, this places a rather strong requirement on the 
quantum error correction code
since constraints on {\em both} bit and phase error rates have to be satisfied. Fortunately,
Lo~\cite{Lo_decouple_2003} further shows that bit and phase errors can be decoupled by simply encrypting the bit error syndrome transmission (without affecting the net key generation rate).
Koashi~\cite{Koashi_Uncer_06} adopts the same decoupling mechanism and further generalizes the notion of phase errors with a simple and yet powerful argument.
In this paper, we follow this line of security proofs in our finite-key analysis.
Essentially, the important ingredients in our analysis are
\begin{itemize}
\item encrypting the bit error syndromes;
\item using a random sampling argument to place bounds on the phase error rate;
\item using a privacy amplification scheme (with structure) and placing bounds on its phase error correcting capacity;
\item integrating authentication in classical communications.
\end{itemize}

\subsection{Composable security}

The finite key analysis is closely related to the definition of security. Currently, the composable security definition of QKD \cite{BenOr:Security:05,Renner:Security:05} is widely accepted as the most stringent security definition in the field.
QKD is composable in the sense that the final key generated is
indistinguishable from an ideal secret key except with a small probability, and thus the key can be used in a subsequent cryptographic task where an ideal key is expected.
A secret key is considered ideal if it is identical between Alice and Bob and is private to Eve (i.e., Eve has no information on it).
The notion of composability was first proposed 
in the classical setting for
the study of security when composing classical cryptographic protocols in a complex manner~\cite{Canetti2001,Canetti2002}.
Composability has also been carried over to the quantum setting~\cite{Ben-Or2004,BenOr:Security:05}.
One essential feature of the composable security definition is that it characterizes the security of a protocol with respect to the ideal functionality.
In particular, the security of a composable secret key generated by QKD is measured with 
the trace distance between the real situation with the real key and the ideal situation with an ideal key%
~\cite{Konig2007,Renner:Security:05,Renner_Thesis_05}.
\begin{definition}[\cite{Konig2007,Renner:Security:05,Renner_Thesis_05}]
\label{def-composability}
{\rm
A random variable $V$ (the classical key) drawn from the set $\mathcal{V}$ is said to be $\zeta$-secure with respect to an eavesdropper holding a quantum system $E$ if
\begin{align}
\label{eqn-composability-def}
\frac{1}{2} \operatorname{Tr} | \rho_{VE} - \rho_{U} \otimes \rho_{E} | \leq \zeta
\end{align}
where 
$\rho_{VE}=\sum_{v \in \mathcal{V}} P_V(v) \ket{v}\bra{v} \otimes \rho_{E|V=v}$, 
$\rho_{U}=\sum_{v \in \mathcal{V}}  \ket{v}\bra{v} / |\mathcal{V}|$
represents an ideal key taking values uniformly over $\mathcal{V}$,
and $|\mathcal{V}|$ is the size of $\mathcal{V}$.
Here, $\operatorname{Tr} |A|=\sum_i |\lambda_i|$ where $\lambda_i$ are the eigenvalues of $A$.
}
\end{definition}
Since the trace distance $\frac{1}{2} \operatorname{Tr}| \rho - \sigma |$ is the maximum probability of distinguishing between the two quantum states $\rho$ and $\sigma$, this security definition naturally gives rise to the operational meaning that the $\zeta$-secure key $V$ is identical to an ideal key $U$ except with probability $\zeta$.
This trace-distance security parameter is additive when practical cryptographic protocols are composed~\cite{BenOr:Security:05}.
That is to say, suppose we have a key generation protocol (e.g., QKD)
and a second cryptographic protocol which consumes an ideal secret key.
Furthermore, we suppose that the first protocol
 realizes an ideal key generation scheme with a security parameter $\zeta_1$ for a particular secret key output, and  also the second protocol 
  realizes its ideal functionality with a security parameter $\zeta_2$.
Then, when the two protocols are composed (i.e., the imperfect key generated in the first protocol is used in the second protocol), the overall security parameter will become $\zeta_1+\zeta_2$.

Our paper is based on EDP-based proofs which often justify security with the fidelity between Alice and Bob's state and the ideal state 
(the perfect EPR pairs).
This fidelity is a direct consequence of the failure probability of the post-processing procedure.
Thus, we need to find a connection of this failure probability with the composability definition in Definition~\ref{def-composability}.

The generation of the final key in one round of QKD is composed of many steps (cf. Sec.~\ref{Sc:Outline} and Table~\ref{table:resources}) and each step carries a certain failure probability.
This probability represents the case that Alice and Bob believe the step has succeeded but actually not (in other word, a case of undetected failure).
A detected failure in any step will lead a premature termination of the QKD process.
Successes of all steps will result in a perfect final key that is private to Eve and identical between Alice and Bob.
However, since each step may fail without being detected, there is a certain probability that the final key fails to be perfect and this probability is upper bounded by the sum of the failure probabilities of all the steps.
Essentially, this sum is the failure probability $\varepsilon$ of the entire post-processing procedure, which needs to be converted to the security parameter of the final key.

In the context of Koashi's proof~\cite{Koashi_Uncer_06}, success of the 
phase error correcting step as part of the post-processing procedure guarantees that
Alice's $m$-qubit state $\rho_A$ can be 
corrected to the pure state $0_X^{\otimes m}$.
The final key is then generated by $m$ measurements in the $Z$-basis on $\rho_A$.
Since the entire post-processing procedure fails with probability at most $\varepsilon$, 
the component of Alice's state $\rho_A$ corresponding to the pure state $0_X^{\otimes m}$ must satisfy
$\langle 0_X^{\otimes m} | \rho_A | 0_X^{\otimes m}\rangle \geq 1-\varepsilon$.
In order to make the connection with the universal composability definition in 
Definition~\ref{def-composability},
it has been suggested~\cite{Konig2007} (also see~\cite{BenOr:Security:05}) that a bound on the trace distance is obtained from the fidelity using a general inequality relating them~\cite{Fuchs1999}:
\begin{align}
\label{eqn-ineq-tracedistance-fidelity}
\frac{1}{2} \operatorname{Tr}| \rho - \sigma | \leq \sqrt{1-F(\rho,\sigma)^2},
\end{align}
where $F(\rho,\sigma)=\operatorname{Tr} \sqrt{\rho^{1/2} \sigma \rho^{1/2}}$ is the fidelity between $\rho$ and $\sigma$.
Thus, we seek the minimum of the fidelity between $\rho_{AE}$ and $\ket{0_X^{\otimes m}}_A \bra{0_X^{\otimes m}} \otimes \rho_E$ in order to get an upper bound on their trace distance, in accordance with Definition~\ref{def-composability}.
Since the fidelity never decreases under a trace-preserving quantum operation (i.e., $F(\mathcal{E}(\rho),\mathcal{E}(\sigma))\geq F(\rho,\sigma)$), system $E$ can be considered to be the entire purification of system $A$ when the minimum occurs.
Assuming this worst case,
the joint state is of the form
\begin{align}
\ket{\Psi_{AE}} = \sqrt{\alpha} \ket{0_X^{\otimes m}}_A \ket{0}_E + \sqrt{1-\alpha} \ket{\Psi^\perp}_{AE}
\end{align}
where $\ket{\Psi^\perp}_{AE}$ has unit norm,  
$_A\langle 0_X^{\otimes m}\ket{\Psi^\perp}_{AE}=0$, and
$\alpha \geq 1-\varepsilon$.
The fidelity between the real situation and the ideal situation is (see Appendix~\ref{App:proof-fidelity-bound} for proof)
\begin{align}
F\left(\rho_{AE},\ket{0_X^{\otimes m}}_A \bra{0_X^{\otimes m}} \otimes \rho_E\right) & \geq \alpha \label{eqn-fidelity-bound-general1} \\
& \geq 1-\varepsilon
\label{eqn-fidelity-bound-general2}
\end{align}
where $\rho_{AE}=\ket{\Psi_{AE}}\bra{\Psi_{AE}}$ and $\rho_E=\operatorname{Tr}_A (\rho_{AE})$.
Thus, an upper bound on the failure probability provided by the EDP-based proofs can easily be translated to a composable security measure.
By substituting Eq.~\eqref{eqn-fidelity-bound-general2} for the fidelity in Eq.~\eqref{eqn-ineq-tracedistance-fidelity} and using the fact that projection onto the eigenstates of the $Z$-basis corresponding to the final measurement does not increase the trace distance~\footnote{The trace distance never increases under a trace-preserving quantum operation, i.e., $\operatorname{Tr}| \mathcal{E}(\rho) - \mathcal{E}(\sigma) | \leq \operatorname{Tr}| \rho - \sigma |$.}, we conclude that 
the final key is $\sqrt{\varepsilon(2-\varepsilon)}$-secure
in accordance with Definition~\ref{def-composability}.

\begin{lemma}
\label{lemma-relation-failure-prob-trace-distance}
{\rm
When the failure probability of the post-processing procedure is $\varepsilon$, the final key is $\sqrt{\varepsilon(2-\varepsilon)}$-secure
in accordance with Definition~\ref{def-composability}.
}
\end{lemma}

We can apply this lemma 
when many rounds of QKD are composed.
Suppose the post-processing of each round fails with a probability $\varepsilon$, and Alice and Bob plan to use a QKD system 
a million
times in the manner that the secret key output of one round is fed as input to the next.
Since the trace-distance measure is additive when cryptographic protocols are composed~\cite{BenOr:Security:05},
the trace-distance security parameter for the key in the last round will be $10^6 \sqrt{\varepsilon(2-\varepsilon)}$. 
Note that this security parameter increases linearly with the number of rounds of the QKD system. This linear dependence is an important feature of the composability security definition.

We note that Mayers' security proof~\cite{Mayers1996,Mayers_01} has also implicitly mentioned using failure probability to quantify the security.

\subsection{Equivalence of the failure probability and the trace distance as the optimization objective\label{sec-equivalence-failure-prob-trace-distance}}

The failure probability of the post-processing procedure $\varepsilon$ is related to the trace-distance security parameter $\zeta$ by $\zeta=\sqrt{\varepsilon(2-\varepsilon)}$.
Since $\frac{d\zeta}{d\varepsilon} > 0$ and $\frac{d^2\zeta}{d\varepsilon^2} < 0$, 
we have $\zeta^{(1)}>\zeta^{(2)} \Leftrightarrow \varepsilon^{(1)}>\varepsilon^{(2)}$ and there is a one-to-one mapping between these two measures.
Thus, minimizing either $\varepsilon$ or $\zeta$ subject to the same constraints (such as a fixed key length) will produce the same solution.

\subsection{Simple lower bound on failure probability}

It is easy to lower bound the failure probability as a function of the secret-key cost $k_\text{initial}$ by considering that Eve has a $2^{-k_\text{initial}}$ chance of guessing the right initial secret key and thus will be able to launch a man-in-the-middle attack successfully.
Therefore, the failure probability of any post-processing scheme should be at least $2^{-k_\text{initial}}$.
Moreover, the failure probability of our scheme exhibits the same exponential decrease as the lower bound (see the various constituent failure probabilities $\varepsilon$'s listed in Table~\ref{table:resources}).

\section{Outline of post-processing procedure} \label{Sc:Outline}
\begin{figure}
\includegraphics[width=1\columnwidth]{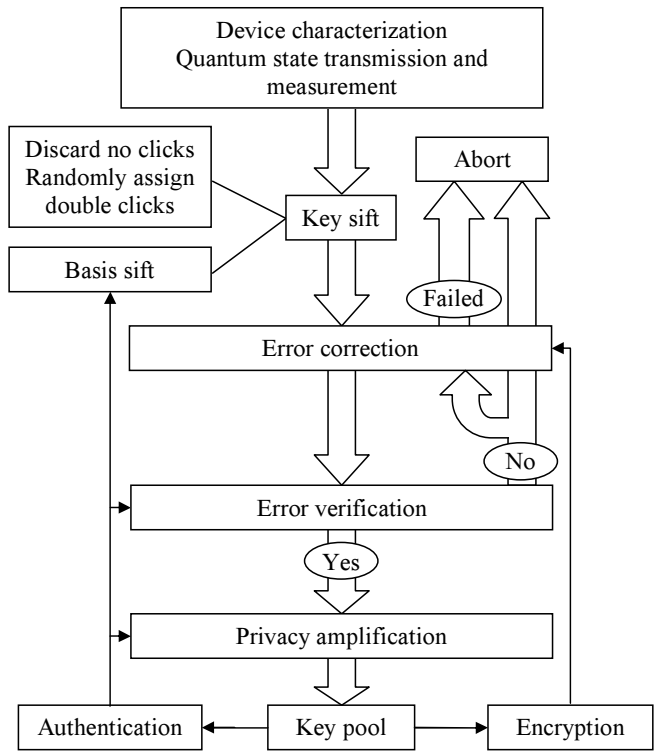}
\caption{\label{fig:flowchart}
Flow chart of our data post-processing procedure~\cite{Tittel2008}.
}
\end{figure}

The post-processing procedure is listed as follows. 
We remark that each communication between Alice and Bob consists of a message and an authentication tag, each of which may be of zero length.
In our scheme, a tag is transmitted if and only if authentication is used and in this case the authentication tag is always encrypted by a one-time pad, consuming some pre-shared secret bits.
When a message is transmitted, it may or may not be encrypted, and it is assumed to be unencrypted unless otherwise stated.
Note that no message but a tag is transmitted in the error verification step (step 4).
Fig.~\ref{fig:flowchart} shows the flow chart of 
our data post-processing procedure.

\begin{enumerate}
\item
Key sift [not authenticated]: 
Alice sends $N$ quantum signals to Bob, of which $n$ signals produce clicks.
Bob discards all no-click events and obtains $n$-bit raw key by randomly assigning values to the double-click events~\footnote{In the case of a passive-basis-selection setup, Bob also randomly assigns basis value $X$ or $Z$ for double clicks~\cite{BML_Squash_08}.}.  Note that other key sift procedures might be applied as well (see, for example, Ref.~\cite{EffLoop_08}).

\item
Basis sift [authenticated]: Alice and Bob send each other $n$-bit basis information. In the end of this step, Alice and Bob obtain $n_x$ ($n_z$)-bit sifted key in the $X$($Z$)-basis. Define the bias ratio to be $q_x\equiv n_x/(n_x+n_z)$. Note that this bias ratio is different from the probability the Alice and Bob choose the two bases. Define the probability that Alice and Bob choose the $X$-basis to be $p_x$, then in the long key limit, $q_x=p_x^2/[p_x^2+(1-p_x)^2]$.

\item

Error correction [not authenticated but encrypted%
, Section~\ref{sec-error-correction}]: Alice and Bob perform error correction so that Bob's raw key matches Alice's.
The classical messages exchanged in this process are encrypted.
If error correction fails, Alice and Bob abort the QKD process.

\item
Error verification [Section~\ref{sec-error-verification}]: Alice and Bob want to make sure (with a high probability) that their keys after the error correction step are identical.
If error verification fails, Alice and Bob either go back to the error correction step or abort the QKD process.
We note that the idea of using error verification to replace error testing is proposed by L\"utkenhaus \cite{Lutkenhaus:practical:99}.

\item
Phase error rate estimation [no communication, Section~\ref{Sc:Sample}]: 
Alice and Bob use the bit error rate measured in the $X$($Z$)-basis to infer the phase error rate in the $Z$($X$)-basis.
The uncertainty in bounding the phase error rates are quantified by a random sampling argument.

\item
Privacy amplification [authenticated, Section~\ref{sec-phase-error-correction}]: Alice randomly generates an $(n_x+n_z+l-1)$-bit random bit string and sends to Bob through an authenticated channel. Alice and Bob use this random bit string to generate a Toeplitz matrix. The final key (with a size of $l$) will be the product of this matrix (with a size of $(n_x+n_z)\times l$) and the key string (with a size of $n_x+n_z$).

\item
The final secure key length (net growth) is given by
\begin{equation} \label{Finite:KeyFinal}
\begin{aligned}
NR &\ge l-k_{bs}-k_{ec}-k_{ev}-k_{pa} \\
\end{aligned}
\end{equation}
with a failure probability of
\begin{equation} \label{Finite:FailFinal}
\begin{aligned}
\varepsilon\le\varepsilon_{bs}+\varepsilon_{ev}+\varepsilon_{ph}+\varepsilon_{pa} \\
\end{aligned}
\end{equation}
where $l$ is given by Eq.~\eqref{Finite:PAkey}.
Here, the $k$'s are the secret-key costs and the $\varepsilon$'s are the failure probabilities for steps 2-6 (see Table~\ref{table:resources}).
Throughout the paper, $\varepsilon$'s with various footnotes stand for various failure probabilities.

\end{enumerate}

\begin{table*}
\begin{tabular}{ |l|c|c|c|c| }
\hline
Step & message length & message encrypted? & tag length & failure probability \\
\hline
1. Key sift & $N$ & - & - & - \\
\hline
2. Basis sift & $2n$ & No & $2k_{bs}$ & $2\varepsilon_{bs}$ [Eq.~\eqref{Finite:ReconFail}] \\
\hline
3. Bit error correction & $k_{ec}$ [Eq.~\eqref{Finite:ECcost}] & Yes & - & -  \\
\hline
4. Error verification & - & - & $k_{ev}$ & $\varepsilon_{ev}$ [Eq.~\eqref{Finite:EVFail}] \\
\hline
5. Phase error estimation & - & - & - & $\varepsilon_{ph}$ [Eq.~\eqref{Finite:PhFail}] \\
\hline
6. Privacy amplification & $(n_x+n_z+l-1)$ & No & $k_{pa}$ & $\varepsilon_{pa}$ [Eq.~\eqref{Finite:PAFail}] \\
\hline
\end{tabular}
\caption{\label{table:resources}List of resource cost and the failure probabilities in the various steps.
Lengths of pre-shared secret key bits are designated with $k$ while the failure probabilities with $\varepsilon$. The relevant equations involving these quantities are also shown.}
\end{table*}

\section{Preliminary\label{sec-preliminary}}

\subsection{Data representation\label{sec-data-representation}}

Data are represented as matrices or column vectors of 0's and 1's.
Additions are carried out in modulo 2.
For example, a raw key $x$ is multiplied by a privacy amplification matrix $M$ to generate the final $y=M x$, where the $i$th bit of $y$ is $\sum_{j} M(i,j) x (j) \mod 2$.

\subsection{Toeplitz matrices}

In our framework, we rely heavily on a particular class of hash functions to perform
phase error correction, error verification, and authentication.
We are interested in using sets of Toeplitz matrices to perform these tasks.
Toeplitz matrices are Boolean matrices with a special structure:
\begin{align}
M=
\begin{pmatrix}
a_0 & a_{-1} & a_{-2} & \cdots & a_{-m+1}\\
a_1 & a_0 & a_{-1} & \ddots \\
a_2 & a_1 & \ddots && \vdots\\
\vdots & \ddots \\
a_{l-1} & & \cdots & & a_{l-m}
\end{pmatrix}
\end{align}
where $a_i=0,1$.
It can also be concisely described by
 $M_{(i,j)}=a_{i-j}$ where $M_{(i,j)}$ is the $(i,j)$ element of $M$.
The advantage of using Toeplitz matrices is that it can be specified by a small number of parameters, namely, $m+l-1$ bits, as opposed to $m l$ bits for completely random matrices.
Hashing of a given column vector $x$ (whose elements are $0$ or $1$) can be performed 
by choosing a Toeplitz matrix $M$ randomly and computing the hash value as $M x$.

In our post-processing scheme,
we use Toeplitz matrices for three purposes: privacy amplification, error verification, and authentication.
We remark that fully random Toeplitz matrices (specified by $m+l-1$ random bits) are used for privacy amplification, while for error verification and authentication, Toeplitz matrices specified by a smaller number ($2l$) of random bits are used in order to save secret bits (see Sec.~\ref{Sc:Auth} below).

\subsection{Authentication%
} \label{Sc:Auth}

Alice and Bob can authenticate their classical communications with a family of Toeplitz matrices.
For every (classical) message they need to authenticate, both parties select a hash function 
from a fixed family using pre-shared secret bits.
The sending party computes the hash value for the message (called the tag) and sends both to the other party, who also computes the hash value for the received message and can conclude that the message originates from the legitimate party
if both hash values are identical.

In authentication,
the key component is the construction of hashing function.
Wegman and Carter proposed unconditional secure authentication schemes \cite{WC_Authen_79,WC_Authen_81} by introducing the universal hashing function families,
which are also used for privacy amplification.
Afterwards, lots of efforts have been spent on how to construct a universal hashing function family effectively. In this paper, we use the LFSR-based Toeplitz matrix construction by 
Krawczyk \cite{Krawczyk:Hash:94,Krawczyk:Hash:95} for authentication.

Here we state the result of the LFSR-based Toeplitz matrix construction, which is
given by Theorem 9 of Ref.~\cite{Krawczyk:Hash:94} by Krawczyk.
The authentication scheme based on the LFSR-based Toeplitz matrix construction is secure with a failure probability of
\begin{equation} \label{Authen:Thm9}
\begin{aligned}
\varepsilon_{au} = 
n 2^{-k+1}
\end{aligned}
\end{equation}
where $k$ is the length of the tag, $n$ is the length of the message. The authentication scheme can be stated as follows. Alice and Bob use a $2k$-bit secure key to construct a Toeplitz matrix with a size of ($k\times n$) by a LFSR. The authenticated tag is generated by multiplying the matrix and the message. Then they encrypt the tag by another $k$-bit secure key. Since the tag is encrypted by a one-time pad, the $2k$-bit key used for the Toeplitz matrix construction is still secure \cite{Krawczyk:Hash:94}. Hence, the net secure-key cost for this scheme is $k$.

We remark that in Krawczyk's later result \cite{Krawczyk:Hash:95}, the secure key required for the LFSR-based Toeplitz matrix construction can be reduced to an arbitrary number $r$, with sacrifice of failure probability,
\begin{equation} \label{Authen:Thm14}
\begin{aligned}
\varepsilon_{au} = \frac{1}{2^k}+\frac{k+n-1}{2^{r/2}}
\end{aligned} .
\end{equation}
One can see that by choosing $r=2k$, Eq.~\eqref{Authen:Thm9} gives a slightly tighter bound than Eq.~\eqref{Authen:Thm14} for the failure probability. Since the secure-key cost is at least $k$ in this construction due to one-time pad encryption, for simplicity, we use Eq.~\eqref{Authen:Thm9} for authentication and error verification.

We remark that, as pointed out in Ref.~\cite{Krawczyk:Hash:94}, the LFSR-based Toeplitz matrix construction is highly practical in real-life implementation.

\section{Basis sift\label{sec-basis-sift}}

Alice and Bob send each other $n$-bit basis information. Due to the symmetry, we can assume the same failure probability for the two message exchanges~\cite{Krawczyk:Hash:94}:
\begin{equation} \label{Finite:ReconFail}
\begin{aligned}
\varepsilon_{bs} &= n 2^{-k_{bs}+1}
\end{aligned} . 
\end{equation}
Here, Alice and Bob use a $2k_{bs}$-bit secure key to construct a Toeplitz matrix with a size of ($k_{bs} \times n$) by a LFSR. The authenticated tag is generated by multiplying the matrix and the message. Then they encrypt the two tags by two $k_{bs}$-bit secure keys. 
The total secure-key cost in this step is $2k_{bs}$ (for the one-time-pad encryption of the tags) and the corresponding failure probability is $2\varepsilon_{bs}$. Note that when Alice and Bob use a biased basis choice \cite{EffBB84_05}, they can exchange less than $n$-bit classical information for basis sift by data compression. Since the secure-key cost only logarithmically depends on the length of the message, we simply use $n$ for the following discussion. In the end of this step, Alice and Bob obtain $n_x$ ($n_z$)-bit sifted key in the $X$ ($Z$) basis. Define the bias ratio to be $q_x\equiv n_x/(n_x+n_z)$ as in Sec.~\ref{Sc:Outline}.

\section{Error correction\label{sec-error-correction}}

For simplicity of discussion, we assume that Bob tries to correct his raw key to match Alice's. 
This means that we assume no advantage distillation~\cite{TwoWay_03,TwoWay_06}.
Error correction is done by Alice sending parity information of her raw key to Bob encrypted with secret bits from the key pool.
The secure-key cost is given by
\begin{equation} \label{Finite:ECcost}
\begin{aligned}
k_{ec}=n_xf(e_{bx})H(e_{bx})+n_zf(e_{bz})H(e_{bz})
\end{aligned}
\end{equation}
where $f(x)$ is the error correction efficiency, and
\begin{equation}
\begin{aligned}
H(x)=-x\log_2(x)-(1-x)\log_2(1-x)
\end{aligned}
\end{equation}
is the binary entropy function.
In practice, Alice and Bob only need to count the amount of classical communication used in the error correction. That is, the value of $k_{ec}$ can be directly obtained from the post-processing. After the error correction, Alice and Bob count the number of errors in the $X$ ($Z$) basis: $e_{bx}n_x$ ($e_{bz}n_z$).
Note that although we assume encryption of the parity information here (cf. Sec.~\ref{sec-method}), 
it may be avoided by basing the post-processing on other security proofs.
In this case, there may be some restriction on the error correction procedure and more privacy amplification may be required.
In practice, there is an advantage to using error correction without encryption, since if Alice and Bob abort the QKD procedure, no secret bits will be lost due to encryption.

There is no failure probability associated with error correction in our post-processing scheme.
Identity between Alice's and Bob's sifted keys is verified with an error verification step (Sec.~\ref{sec-error-verification} below).

\section{Error verification\label{sec-error-verification}}

Suppose Alice and Bob each holds a bit string $\mathbf{a}$ and $\mathbf{b}$.
They can verify the identity of their strings by exchanging shorter strings which are the hash values $f(\mathbf{a})$ and $f(\mathbf{b})$.
Identity of the two hash values provides confidence that the two strings are the same.
Below we argue that error verification is the same as authentication, and thus we use the same procedure for both purposes.
This procedure and the associated properties is described in the authentication section (Sec.~\ref{Sc:Auth}).

\subsection{Relation to authentication}

In QKD post-processing, authenticated classical communication is required to overcome the man-in-the-middle attack. The objective of the authentication procedure can be stated as follows. Alice sends Bob a message through a (classical) channel, which is accessible to Eve. Alice uses some authentication scheme to make sure (with a high probability) that the message is not modified during the transmission. This classical problem is well studied in the literature \cite{WC_Authen_79,WC_Authen_81,Krawczyk:Hash:94,Krawczyk:Hash:95}. One traditional solution is for Alice to add a redundant code (tag) to the message to be sent. The tag-message pair is designed in such a way that whenever the message is modified, Bob can detect it (with a high probability).

Error verification, on the other hand, is the procedure for ensuring (with a high probability) that the bit strings (or keys) owned by Alice and Bob are identical. One natural way to do this is by random hashing. For example, Alice randomly hashes her bit string and sends the hash value to Bob. Bob uses the same hash function to obtain his hash value and compares to the one sent by Alice. The probability that Alice and Bob possess different bit strings (keys) decrease exponentially with the number of rounds of hashing.

By comparing the two procedures, authentication and error verification, one can see their commonality. In order to show the link between the two procedures, we break down the authentication procedure into two parts: Alice sends Bob the message first and then the tag. Let us take a look at the stage where Bob just receives the message sent by Alice (but before the tag). Now, Alice and Bob each has a bit string. In authentication, Alice sends a tag (corresponding to her message) to Bob and Bob verifies it. The claim of a secure authentication scheme is that if the tag passes Bob's test, the probability that Alice and Bob share the same string is high. This is exactly what is asked in the error verification procedure. Therefore, secure authentication schemes can be used for the error verification.

We remark
that the only difference between the two procedures is that authentication does not care whether the tag reveals information about the message or not, but error verification does (at least for our use in QKD post-processing). This difference can be easily overcome by encrypting the tag, which has already been done in some authentication schemes including the one we use 
in this paper.

Thus, in this procedure, Alice sends an encrypted tag of an authentication scheme to Bob. The failure probability for this step, $k_{ev}$, similar to Eq.~\eqref{Finite:ReconFail}, is
\begin{equation} \label{Finite:EVFail}
\begin{aligned}
\varepsilon_{ev} = (n_x+n_z)2^{-k_{ev}+1}
\end{aligned}.
\end{equation}

\section{Phase error rate estimation} \label{Sc:Sample}
In the BB84 protocol, Alice and Bob measure the bit error rate in the $X$-basis, $e_{bx}$, to estimate the phase error rate in the $Z$-basis, $e_{pz}$, and vice versa. In the infinite length limit, the error {\em rates}, $e_{bx}$ and $e_{pz}$, converge to the underlying probabilities, $p_{bx}$ and $p_{pz}$. Due to the symmetry of BB84, we know that $p_{bx}=p_{pz}$, from which follows $e_{bx}=e_{pz}$ in the asymptotic case. With a finite key size, the rate is fluctuating around the corresponding probability. Now the question can be stated as a random sampling problem: given the bit error rate in the $X$-basis~($e_{bx}$), the sample size~($n_x$), and the population size~($n_x+n_z$), upper bound the phase error rate in the $Z$-basis~($e_{pz}$), with a probability $1-P_{\theta x}$,
\begin{equation} \label{Random:ConfiDef}
\begin{aligned}
P_{\theta x} &\equiv \textnormal{Pr}\{e_{pz} \ge e_{bx}+\theta_x\}, \\
\end{aligned}
\end{equation}
where $\theta_x$ represents the deviation of the phase error rate from the tested value (the bit error rate in another basis) due to the finite-size effect. 
Here $n_x$ and $n_z$ are the number of sifted key bits in the $X$- and $Z$-basis, respectively. The failure probability $P_{\theta x}$ will be related to the failure probability of 
the phase error rate estimation step (see Eq.~\eqref{Finite:PhFail}).

\subsection{Random sampling}
Define two random variables: $k\equiv e_{bx}n_x$ and $m\equiv e_{pz}n_z+e_{bx}n_x$. The number of bit errors in the $X$-basis, $k$, can be accurately (with a high probability) counted after the error verification procedure. Note that $m$ denotes the number of bit errors \emph{if} Bob measures all $n_x+n_z$ qubits in the $X$-basis. In the squashing model~\cite{GLLP_04,TT_Thres_08,BML_Squash_08}, Eve prepares the qubit received by Bob. Hence, one can assume Eve chooses a distribution of $m$, $\textnormal{Pr}\{m\}$, before Bob's detection.

In order to link the probability, $P_{\theta x}$, to the measurement results, $n_x$, $n_z$ and $k$ (or $e_{bx}$), we go back to the original definition of the security parameter in QKD. In the security analysis of QKD, $P_{\theta x}$ denotes the probability that Eve sets up a distribution $\textnormal{Pr}\{m\}$ (by preparing qubits) and then Bob obtains $k$ bit errors in the $X$-basis. Therefore, the mathematical definition for $P_{\theta x}$ is
\begin{equation} \label{Random:Ptheta}
\begin{aligned}
P_{\theta x} &= \textnormal{Pr}\{ e_{pz} \ge e_{bx}+\theta_x,e_{bx} \} \\
&= \textnormal{Pr}\{ m \ge e_{bx}(n_x+n_z)+\theta_xn_z,k \} \\
&= \sum_{m=e_{bx}(n_x+n_z)+\theta_xn_z}^{e_{bx}n_x+n_z} \textnormal{Pr}\{ m,k \} \\
&= \sum_{m=e_{bx}(n_x+n_z)+\theta_xn_z}^{e_{bx}n_x+n_z} \textnormal{Pr}\{ k|m \} \textnormal{Pr}\{ m \} . \\
\end{aligned}
\end{equation}
Bob chooses to measure the $X$-basis randomly (without replacement, of course), and thus $\textnormal{Pr}\{ k|m \}$ is given by a hypergeometric function,
\begin{equation} \label{Random:Hypergeo}
\begin{aligned}
\textnormal{Pr}\{ k|m \} = {{{m \choose k} {{n_x+n_z-m} \choose {n_x-k}}}\over {n_x+n_z \choose n_x}}.
\end{aligned}
\end{equation}
It is not hard to prove that Eq.~\eqref{Random:Hypergeo} is a strict decreasing function of $m$ when $m>(n_x+n_z)e_{bx}$. Thus, from Eq.~\eqref{Random:Ptheta},
\begin{equation} \label{Random:Pbound}
\begin{aligned}
P_{\theta x} &\le \textnormal{Pr}\{ k|m=e_{bx}(n_x+n_z)+\theta_xn_z \} \\
&< \frac{\sqrt{n_x+n_z}}{\sqrt{e_{bx}(1-e_{bx})n_xn_z}} 2^{ -(n_x+n_z)\xi_x(\theta_x) } \\
\end{aligned}
\end{equation}
where the first equality holds when Eve sets the probability distribution to be a delta function $\textnormal{Pr}\{ m \}=\delta_{m,e_{bx}(n_x+n_z)+\theta_xn_z}$. The derivation of the second inequality is presented in Appendix~\ref{App:Hyper}. Note that all the variables in Eq.~\eqref{Random:Pbound} can be measured in practice. The function $\xi_x(\theta)$ is given by
\begin{equation} \label{Random:XiPhase}
\begin{aligned}
\xi_x(\theta_x) \equiv & H(e_{bx}+\theta_x-q_x\theta_x)-q_xH(e_{bx})\\
&-(1-q_x)H(e_{bx}+\theta_x) 
\end{aligned}
\end{equation}
where $q_x=n_x/(n_x+n_z)$ is the bias ratio.

A similar formula for the failure probability of phase error rate estimation in the $X$-basis, $P_{\theta z}$, can also be derived,
\begin{equation} \label{Finite:PhFailX}
\begin{aligned}
P_{\theta z} &<
\frac{\sqrt{n_x+n_z}}{\sqrt{e_{bz}(1-e_{bz})n_xn_z}} 2^{ -(n_x+n_z)\xi_z(\theta_z) } \\
\end{aligned},
\end{equation}
with $\xi_z(\theta_z)$ is defined by $\xi_z(\theta_z) \equiv H(e_{bz}+\theta_z-q_z\theta_z)-q_zH(e_{bz})-(1-q_z)H(e_{bz}+\theta_z)$ and $q_z=n_z/(n_x+n_z)$.
Combining the failure probabilities for the X-basis and Z-basis,
the total failure probability of phase error rate estimation, $\varepsilon_{ph}$, is then given by
\begin{equation} \label{Finite:PhFail}
\begin{aligned}
\varepsilon_{ph} &\le P_{\theta x}+P_{\theta z}
\end{aligned}.
\end{equation}

In case $e_{bx}=0$ (or $e_{bz}=0$), one can replace it by $n_xe_{bx}=1$ (or $n_ze_{bz}=1$) to get around the singularity as shown in Appendix \ref{App:Hyper}. One can see that $\xi_x(\theta_x)$ is positive when $\theta_x>0$ and $0\le e_{bx}, e_{bx}+\theta_x \le1$, due to concavity of the binary entropy function $H(x)$.

\subsection{Large data size approximation}
In the limit of large $n_x$ and $n_z$, $\theta_x$ can be chosen to be small. Then we can use Taylor expansion for Eq.~\eqref{Random:XiPhase},
\begin{equation} \label{Random:XiTheta}
\begin{aligned}
\xi_x(\theta_x) =& H(e_{bx}+\theta_x-q_x\theta_x)-q_xH(e_{bx})\\
&-(1-q_x)H(e_{bx}+\theta) \\
=& -\frac12(1-q_x)q_xH''(e_{bx})\theta_x^2+O(\theta_x^3) \\
=& \frac{\ln2}2\frac{(1-q_x)q_x}{(1-e_{bx})e_{bx}}\theta_x^2+O(\theta_x^3) .
\end{aligned}
\end{equation}
When $q_x=1/2$, i.e., $n_x=n_z$, and $\theta_x$ is small, the failure probability is given by
\begin{equation} \label{Random:FailTheta}
\begin{aligned}
P_{\theta_x} &< \frac{1}{2\sqrt{2ne_{bx}(1-e_{bx})}} e^{ -\frac{\theta_x^2n}{4(1-e_{bx})e_{bx}} } \\
\end{aligned} .
\end{equation}
Except for the factor 
$1/\left[2\sqrt{2ne_{bx}(1-e_{bx})}\right]$, this is what is used in the literature, such as Refs.~\cite{ShorPreskill_00,EntanglementPDC_07}. In practice, normally we have $2\sqrt{2ne_{bx}(1-e_{bx})}>1$, so the bound given by Eq.~\eqref{Random:FailTheta} is tighter than what is used in the literature.

\section{Privacy amplification\label{sec-phase-error-correction}}

In view of the EDP picture, we regard privacy amplification as the result of phase error correction.
In the following, we focus on using two-universal hashing to perform phase error correction and determine the corresponding failure probability.

\subsection{Two-universal hashing}

The family of all Toeplitz matrices $\{M\}$ of size $l \times m$ has $2^{l+m-1}$ elements and satisfies the following property:
\begin{align}
\label{eqn-Toeplitz-property1}
\text{Pr} \{ M x = M y\} = \frac{1}{2^{l}} \hspace{1cm} \text{for all } x \neq y,
\end{align}
where it is assumed that each matrix is chosen with equal probability.
This can be proved by slightly adapting the proof of Claim 7 of Ref.~\cite{Mansour1990}.
Note that the family of hash functions performed with Toeplitz matrices is one specific case of a more general class known as the two-universal families of hash functions.
A family of hash functions 
mapping $S$ to $T$ is called two-universal~\cite{WC_Authen_79} if
\begin{align}
\label{eqn-2universal-def1}
\text{Pr} \{ f(x) = f(y)\} \leq \frac{1}{|T|} \hspace{1cm} \text{for all } x \neq y,
\end{align}
where $f(x)$ is a hash function chosen in 
the family of hash functions 
and in our case $f(x)=M x$. 
Two-universal families of hash functions have many useful properties and we will rely on some of the them in this paper.

\subsection{Error correction\label{sec-PA-EC}}

Suppose Alice holds a bit string, $\mathbf{a}$, and Bob a noisy version of it, $\mathbf{b}$.
The difference between the two strings $\mathbf{e}=\mathbf{a}\oplus\mathbf{b}$ is the error pattern.
Let $S$ be the set of all possible error patterns.
Alice and Bob intend to use a family of
two-universal
linear hash functions to correct errors in Bob's string with respect to Alice's.
A hash function $f(\cdot)$ is selected from the family and Alice and Bob each computes the hash value of their bit strings with the hash function.
Alice sends Bob her hash value, to which Bob adds his hash value to arrive at the hash value of the error pattern $f(\mathbf{e})=f(\mathbf{a})\oplus f(\mathbf{b})$.
Note that this is valid due to the linearity of the hash functions.
Using this hash value, Bob can identify the error pattern and thus correct the errors in his string.
Suppose that there are
$|T|$
possible outputs for this family of hash functions.
Using the definition of a two-universal family given in Eq.~\eqref{eqn-2universal-def1},
we can
bound the probability of incorrectly identifying the error pattern as
\begin{align}
\label{eqn-error-correction1}
\text{Pr} \left\{ \bigcup_{\mathbf{e'} \in S \setminus \mathbf{e}} f(\mathbf{e}) = f(\mathbf{e'}) \right\}
\leq \frac{|S|}{|T|}
\end{align}
which follows from applying the union bound to
Eq.~\eqref{eqn-2universal-def1}.
Thus, Bob's error-corrected string matches Alice's with probability at least
$1-|S|/|T|$.

Although this hashing-based error correcting procedure may not be as practical and efficient as conventional ones, it is useful for
phase error correction 
in security proofs~\cite{ShorPreskill_00,Lo_decouple_2003,Koashi_Uncer_06}.
This is because for security purpose we only need to show that the phase error pattern is identified without actually correcting the error~\cite{ShorPreskill_00}, and
we only need a bound on the probability of successfully identifying the error pattern.

\subsection{Privacy amplification and phase error correction} \label{Sc:PrAm}

Suppose we perform privacy amplification using a set of $l \times m$ Toeplitz matrices, a member of which can be selected with $l+m-1$ random bits.
Here, $l$ is the final key length and $m$ is the sifted key length.
For each matrix $M$ in the privacy amplification set, we associate an $(m-l) \times m$ matrix $M^\perp$ that is orthogonal to $M$.
The collection of all these matrices $\{M^\perp\}$ forms the set of hash functions for phase error correction.
We would like to find out whether this set $\{M^\perp\}$ has the property of
Eq.~\eqref{eqn-2universal-def1}.
If it does, we can determine the successful probability of phase error correction from Eq.~\eqref{eqn-error-correction1}.

We remark that it does not matter whether the matrices of the set $\{M^\perp\}$ have the Toeplitz form or not since we do not need to generate them but only need to make sure that there exists such a set with a certain phase error correcting capability.
On the other hand, we do impose the Toeplitz form on the privacy amplification set $\{M\}$ since we actually need to generate this set.

Indeed, it can be shown (see, e.g., Theorem 1 of Ref.~\cite{Hayashi:Teo:07}) that when $M$ is chosen from a set of random Toeplitz matrices, the corresponding matrices $M^\perp$ also form a two-universal set, i.e.,
\begin{align*}
\text{Pr} \{ M^\perp x = M^\perp y\} \leq \frac{1}{2^{m-l}} \hspace{1cm} \text{for all } x \neq y .
\end{align*}
Thus, according to the discussion in Sec.~\ref{sec-PA-EC}, we can use the set $\{M^\perp\}$ to identify the phase error pattern and perform the correction.
In essence, when (i) there are $|S|$ number of possible phase error patterns, (ii) the sifted key length is $m$, and (iii) the final key length is $l$,
the failure probability of
phase error correction
is upper bounded by
\begin{align}
\label{eqn-phase-error-correction-failprob}
\varepsilon_{pc} =
|S| 2^{-(m-l)}.
\end{align}
(Note that the sifted key length here will be equal to 
$m=n_x+n_z$ 
when it is used in the next subsection.)

For BB84, the bit error rates for the $Z$ bits and $X$ bits are exactly determined from the error correction procedures, up to a certain probability given by the verification step (cf. Eq.~\eqref{Finite:EVFail}).
Focusing on the $Z$ bits,
we can estimate its phase error rate 
$e_{pz}$
from the actual bit error rate of the $X$ bits $e_{bx}$ using the random sampling argument of Sec.~\ref{Sc:Sample}.
Accordingly,
the lower and upper bounds on the number of phase errors on the $Z$ bits are, except with probability $P_{\theta_x}$ (which is bounded in Eq.~\eqref{Random:Pbound}),
\begin{align}
0 \leq e_{pz} n_z < (e_{bx}+\theta_x) n_z .
\end{align}
(Note that the second inequality is a strictly less than due to the definition of $P_{\theta_x}$ in Eq.~\eqref{Random:ConfiDef}.)
Therefore, with probability at least $1-P_{\theta_x}$, the number of possible phase error patterns in the $Z$ bits is
\begin{align}
|S_z| &= \sum_{k= 0}^{\lceil  (e_{bx}+\theta_x) n_z -1 \rceil}
\binom{n_z}{k} \nonumber \\
&< \binom{n_z}{(e_{bx}+\theta_x) n_z} \label{eqn-number-of-phase-errors-binomial}\\
&< 2^{n_z H(e_{bx}+\theta_x)} , \label{eqn-number-of-phase-errors}
\end{align}
where the first inequality holds for $e_{bx}+\theta_x<1/3$ (see Appendix~\ref{app-claim-binomial-sum} for proof of the first inequality and see, e.g., Refs.~\cite{Feller1968,Cover2006} for proof of the second inequality).
We can similarly obtain the bound for the number of possible phase error patterns in the $X$ bits $|S_x|$.
Combining the number of patterns in the $Z$ and $X$ bits, we have
$|S|=|S_z||S_x|$ in Eq.~\eqref{eqn-phase-error-correction-failprob}.

\subsection{Key length}
Alice and Bob determine the size of the matrix, $l\times (n_x+n_z)$, used for hashing. Here, $l$ is the key length after the privacy amplification. Alice generates an $(n_x+n_z+l-1)$-bit random bit string and sends it to Bob through an authenticated channel. Alice and Bob use this random bit string to generate a Toeplitz matrix. The final key (with a size of $l$) will be the product of this matrix (with a size of $l\times (n_x+n_z)$) and the key string (with a size of $n_x+n_z$) after passing through the error verification. 
The overall failure probability of the privacy amplification is
the sum of that for authentication (Eq.~\eqref{Authen:Thm9}) and that for phase error correction (Eq.~\eqref{eqn-phase-error-correction-failprob}):
\begin{equation} \label{Finite:PAFail}
\begin{aligned}
\varepsilon_{pa} &= (n_x+n_z+l-1) 2^{-k_{pa}+1} + 2^{-t_{oe}}
\end{aligned}
\end{equation}
where $k_{pa}$ is the secure-key cost for the authentication and $t_{oe}$ is related to Eq.~\eqref{eqn-phase-error-correction-failprob} by $2^{-t_{oe}}=\varepsilon_{pc}$.
By rearranging Eqs.~\eqref{eqn-phase-error-correction-failprob}
and \eqref{eqn-number-of-phase-errors}, the final key length is
\begin{equation} \label{Finite:PAkey}
\begin{aligned}
l &= n_x[1-H(e_{bz}+\theta_z)] \\
&~~~+n_z[1-H(e_{bx}+\theta_x)]-t_{oe} . \\
\end{aligned}
\end{equation}
The first term in Eq.~\eqref{Finite:PAFail} gives the failure probability of the authentication for the $(n_x+n_z+l-1)$-bit random bit string transmission. The second term in Eq.~\eqref{Finite:PAFail} gives the failure probability of privacy amplification using the Toeplitz matrix. In the equivalent EDP
used in the security proof \cite{ShorPreskill_00,Koashi_Uncer_06}, the second term in Eq.~\eqref{Finite:PAFail} gives the failure probability of the phase error correction.

\section{Optimization} \label{Sc:Optim}
Alice and Bob calibrate the QKD system to get an estimate of the transmittance $\eta$, the error rates $e_{bx}$ and $e_{bz}$. Through some rough calculation of the target length of the final key, they decide the acceptable confidence interval $1-\varepsilon$ and fix the length of the experiment, $N$, which denotes the number pulses sent by Alice. Then roughly, the length of the raw key is $n=N\eta$. After basis sift, Alice and Bob share an $n_x$-bit ($n_z$-bit) key in the $X$ ($Z$)-basis.

Alice and Bob can optimize their post-processing using either the failure probability or the trace distance as the security measure, since they are directly related to one another as discussed in Sec.~\ref{sec-equivalence-failure-prob-trace-distance}.
Here, we will use the failure probability as the security measure for our discussion.
The failure probability $\varepsilon$ is chosen by Alice and Bob according to the later practical use of the final key. 
The desired message security level sets an upper bound threshold value for 
$\varepsilon$.
Thus, the exact value of $\varepsilon$ is not strictly pre-determined. 
That is, 
it can slightly deviate from the pre-determined threshold value.

Given $n$ and $\varepsilon$ (cf. Eq.~\eqref{Finite:FailFinal}), Alice and Bob need to optimize all the parameters for the key generation rate given in Eq.~\eqref{Finite:KeyFinal}. The first parameter they want to optimize is the basis bias ratio%
, $q_x=n_x/(n_x+n_z)$ which (roughly) determines the probabilities to choose the $X$ and $Z$ bases, $p_x$ and $p_z$, by $q_x\approx p_x^2/(p_x^2+p_z^2)$.
The bias ratio should be determined before quantum transmission while all other parameters can be determined right after a raw key is obtained.
The initial calibration process gives Alice and Bob some idea about the basis ratio which they will use in the subsequent QKD process.
The remaining parameters that need to be optimized are as follows: $k_{bs}$, $k_{ec}$, $k_{ev}$, $k_{pa}$, $\varepsilon_{bs}$, $\varepsilon_{ev}$, $\varepsilon_{ph}$, $\varepsilon_{pa}$ and $t_{oe}$. Alice and Bob need to balance the failure probabilities from each step (cf. Eq.~\eqref{Finite:FailFinal}) and the secure-key cost (cf. Eq.~\eqref{Finite:KeyFinal}).
The optimization problem becomes the following:
given the total failure probability
\begin{equation} 
\begin{aligned}
\varepsilon \le & 2\varepsilon_{bs}+\varepsilon_{ev}+\varepsilon_{ph}+\varepsilon_{pa} \\
=& 2n 2^{-k_{bs}+1} + (n_x+n_z)2^{-k_{ev}+1} + \varepsilon_{ph}\\
&+  (n_x+n_z+l-1) 2^{-k_{pa}+1} + 2^{-t_{oe}}, 
\end{aligned}
\end{equation}
maximize the final key length
\begin{equation} 
\begin{aligned}
NR &\ge l-2k_{bs}-k_{ec}-k_{ev}-k_{pa} \\
\end{aligned}.
\end{equation}

Note that the parameters $k_{bs}$, $k_{ev}$, $k_{pa}$ and $t_{oe}$ affect $\varepsilon$ and the final key rate in similar ways. Also, error correction and phase error rate estimation mainly depend on the bias ratio. Thus, Alice and Bob can group the secure key costs and failure probabilities into two parts
by defining $\varepsilon_3\equiv2\varepsilon_{bs}+\varepsilon_{ev}+\varepsilon_{pa}$ and $k_3\equiv 2k_{bs}+k_{ev}+k_{pa}+t_{oe}$ (see Eqs.~\eqref{Finite:PAkey}, \eqref{Finite:KeyFinal}, and \eqref{Finite:FailFinal}). The final secure key length can be rewritten as
\begin{equation} \label{Finite:KeyOpt}
\begin{aligned}
NR &\ge n_x[1-f(e_{bx})H(e_{bx})-H(e_{bz}+\theta_z)] \\
&~~~+n_z[1-f(e_{bz})H(e_{bz})-H(e_{bx}+\theta_x)]-k_3.
\end{aligned}
\end{equation}
We remark that if the contribution from one basis is negative in Eq.~\eqref{Finite:KeyOpt}, Alice and Bob should use the detections from this basis for parameter estimation only, but not for the key.

We consider the subproblem: given the failure probability
\begin{equation} 
\begin{aligned}
\varepsilon_3 \le & 2\varepsilon_{bs}+\varepsilon_{ev}+\varepsilon_{pa} \\
=& 2n 2^{-k_{bs}+1} + (n_x+n_z)2^{-k_{ev}+1} \\
&+  (n_x+n_z+l-1) 2^{-k_{pa}+1} + 2^{-t_{oe}}, 
\end{aligned}
\end{equation}
minimize the secret-key cost
\begin{equation} 
\begin{aligned}
k_3 &\ge 2k_{bs}+k_{ev}+k_{pa} \\
\end{aligned}.
\end{equation}
With the inequality of arithmetic and geometric means, one can show that
the optimized secure-key cost for each step is given by
\begin{equation} \label{Opt:Opt3k}
\begin{aligned}
t_{oe} &= \frac{k_3}{5}-\frac45-\frac1{5}\log_2A \\
k_{bs} 
&= t_{oe}+1+\log_2n \\
k_{ev} 
&= t_{oe}+1+\log_2(n_x+n_z) \\
k_{pa} 
&= t_{oe}+1+\log_2(n_x+n_z+l-1) ,\\
\end{aligned}
\end{equation}
where $A=n^2(n_x+n_z)(n_x+n_z+l-1)$. The corresponding failure probability is
\begin{equation} \label{Finite:Opt3e}
\begin{aligned}
\varepsilon_{3} = 5A^{1/5}2^{-(k_3-4)/5} \\
\end{aligned}.
\end{equation}
From Eq.~\eqref{Finite:Opt3e}, we have
\begin{equation} \label{Opt:k3e}
\begin{aligned}
k_3 = -5\log_2\varepsilon_{3}+\log_2A+4+5\log_25 .
\end{aligned}
\end{equation}
Note that $n^4/4<A<2n^4$ and also $\varepsilon=\varepsilon_{3}+\varepsilon_{ph}$.
Here if Alice and Bob allow $\varepsilon$ to have a small deviation from the pre-determined value, they can put a \emph{soft} lower bound for $\varepsilon_{3}$ in the optimization.
The exact value of the \emph{soft} lower bound is not really important here as long as it is within the tolerable fluctuation range of $\varepsilon$. Here, we simply choose the tolerable deviation to be within $1\%$, which implies that $10^{-2}\varepsilon<\varepsilon_{3}<\varepsilon$.
Thus,
\begin{equation} \label{Opt:k3bound}
\begin{aligned}
-5\log_2\varepsilon+ & 4\log_2n+2+5\log_25 < k_3 \\
& < -5\log_2\varepsilon+4\log_2n+15+15\log_25 .
\end{aligned}
\end{equation}
This is true for all $\theta_x$, $\theta_z$ and $q_x$. Note that the difference between the lower bound and the soft upper bound of $k_3$ is less than 37 bits. When the final key length is much larger than 37 bits, Alice and Bob can set
\begin{equation} \label{Finite:kmax}
\begin{aligned}
k_3 = -5\log_2\varepsilon+4\log_2n+50 \\
\end{aligned}
\end{equation}
and the failure probability $\varepsilon_3$ will satisfy $\varepsilon_3<10^{-2}\varepsilon$ since the right-hand side of Eq.~\eqref{Finite:kmax} is larger than the upper bound in Eq.~\eqref{Opt:k3bound}.

Since Alice and Bob will recalculate the failure probability in the end and allow the final $\varepsilon$ to have a small deviation from the predefined value, they can safely use $\varepsilon_{ph}=\varepsilon$ in the optimization of the basis bias. Thus, the simplified optimization problem only has three parameters to be optimized: $q_x$, $\theta_x$ and $\theta_z$, given $\varepsilon_{ph}=\varepsilon-\varepsilon_{3}\approx\varepsilon$.

In summary, the simplified optimization procedure for a target failure probability $\varepsilon$ is as follows:
\begin{enumerate}
\item
Compute $k_3$ using Eq.~\eqref{Finite:kmax};

\item
Maximize the key rate in Eq.~\eqref{Finite:KeyOpt} over $q_x$, $\theta_x$, and $\theta_z$ subject to $\varepsilon_{ph}=\varepsilon$. Here, $\varepsilon_{ph}$ is related to the three optimization variables by Eqs.~\eqref{Random:Pbound}, \eqref{Finite:PhFailX}, and \eqref{Finite:PhFail};

\item
After the optimization, they can recalculate the final failure probability
$\varepsilon=\varepsilon_3+\varepsilon_{ph}$, where $\varepsilon_3$ is given in Eq.~\eqref{Finite:Opt3e}.
\end{enumerate}

As discussed above, since one can set $\varepsilon_3<10^{-2}\varepsilon$ (when the key length is much larger than 37 bits), the failure probabilities for basis sift, error verification, and privacy amplification are relatively small, and
the failure probability for random sampling is the major contribution to the total failure probability.

\begin{observation} \label{Obs:MainPhase}
{\rm
The main effect of the finite key analysis for the QKD post-processing stems from the phase error rate estimation.
Inefficiencies due to authentication, error verification, and privacy amplification are relatively insignificant.
}
\end{observation}

\section{Simulations} \label{Sc:Simul}
Now let us consider an example of the post-processing in the simple case of symmetric errors in the two bases.

Suppose $N=10^{10}$, $\eta=10^{-3}$, (then $n=N\eta=10^7$), $e_{bx}=e_{bz}=4\%$ and $\varepsilon=10^{-7}$. It is not hard to see that the final key length is much larger than $37$ bits. Thus, the simplified optimization is used.

First, the secure-key cost $k_3=543$ bit, according to Eq.~\eqref{Finite:kmax}.

Second, given $n=10^7$, $e_{bx}=e_{bz}=4\%$ and $\varepsilon=10^{-7}$, we optimize the parameters: $\theta_x$, $\theta_z$ and $q_x$. Through a numerical program, we get $\theta_x=1.07\%$, $\theta_z=0.84\%$ and $q_x=99.8\%$ (or $p_x=96.0\%$).
Note that, in this case, the $X$ and $Z$ bases are interchangeable due to symmetry.

Finally, we can compute the key length and the corresponding security parameter using our post-processing procedure and compare with the key length using asymptotic assumptions.
The final key length using asymptotic assumptions is 
\begin{align}
K_\text{asymp} = n [1-h_2(e_{bx})-h_2(e_{bz})],
\end{align}
where we used the fact that, asymptotically, the phase error rate in one basis is the same as the bit error rate in the other basis and the use of efficient BB84 leads to always matching basis between Alice and Bob.
The key length with asymptotic analysis is $5.15$ Mb, and the one with the post-processing procedure is $4.41$ Mb and its failure probability is $\varepsilon=1.0073\times10^{-7}$ (roughly $1+2^{-7}$ times the predefined value of $10^{-7}$).
Furthermore, we can get the trace-distance security parameter using Lemma~\ref{lemma-relation-failure-prob-trace-distance} to conclude that this $4.41$ Mb key is composable and is ($4.4884\times10^{-4}$) secure in accordance with security Definition~\ref{def-composability}.
Here, for illustrative purposes, the key length using the post-processing procedure is calculated with the assumption that $n_x=n p_x^2$ and $n_z=n (1-p_x)^2$.

In the simulation, we assume the error correction efficiency is 100\% (Shannon limit). Thus, the difference between the ``asymptotic-key" length and the ``finite-key" length, $0.74$~Mb, comes from the finite statistical analysis. The cost (and the security parameter) due to the finite key analysis mainly comes from the phase error rate estimation. Note that all the remaining cost is only $k_3=543$ bit and $\varepsilon_3=7.3\times10^{-10}$. This point can be clearly seen by comparing Eqs.~\eqref{Finite:PhFail} and \eqref{Finite:Opt3e} in the case of large $n$. The exponent coefficient in Eq.~\eqref{Finite:PhFail} is 
$-\frac{\theta^2}{4(1-e_{bx})e_{bx}}$,
while in Eq.~\eqref{Finite:Opt3e} it is 
$-\frac{k_3}{5n}$, and also a small change in $\theta$ affects the key rate more than that in $k_3/n$ does.

\begin{figure}
\includegraphics[width=\columnwidth]{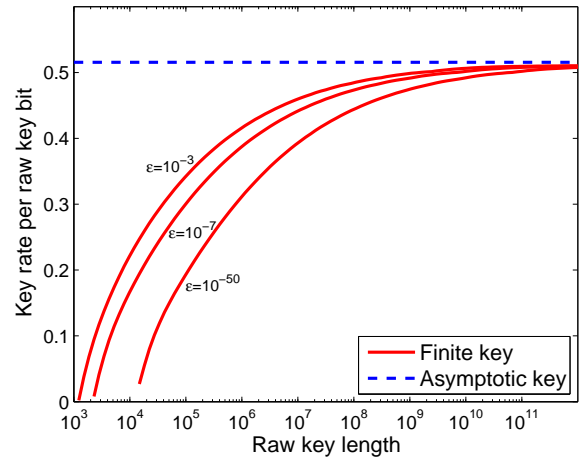}
\caption{\label{fig:RN3}
Lower bound for the key rate as a function of the raw key length; parameters used:
$e_{bx}=e_{bz}=4\%$ 
and
the error correction efficiency is 100\%.
The three curves correspond to three different values of failure probability $\varepsilon$.
}
\end{figure}
Figure~\ref{fig:RN3} shows the lower bound for the key rate as a function of the raw key length.
Note that since we use our simplified optimization method, the final security parameter $\varepsilon$ for the failure probability deviates slightly from the predefined value.
Calculations show that this difference is less than $1\%$ of the predefined values over the entire plotting range for all three curves.

\begin{figure}
\includegraphics[width=\columnwidth]{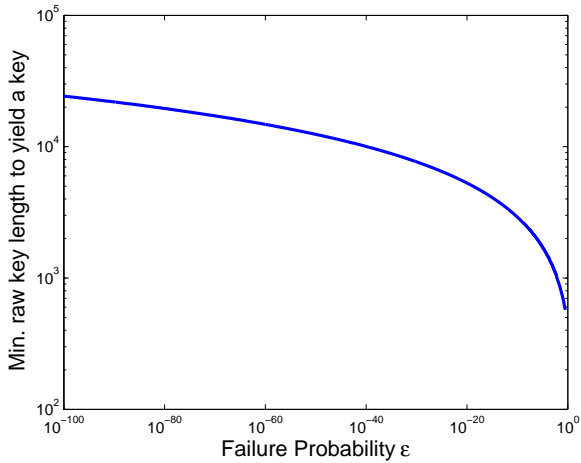}
\caption{\label{fig:Mindatasize}
Minimum 
raw key length
to yield a positive key length as a function of $\varepsilon$; parameters used:
$e_{bx}=e_{bz}=4\%$ and the error correction efficiency is 100\%.
}
\end{figure}
Figure~\ref{fig:Mindatasize} shows the minimum raw key length needed to yield a positive key length as a function of the predefined security parameter $\varepsilon$.
In typical applications, a rough security level may be required for a secret key which is to be generated by QKD.
Thus, this figure gives the minimum number of signals needed to be detected in order to achieve such a security level.

\begin{figure}
\includegraphics[width=\columnwidth]{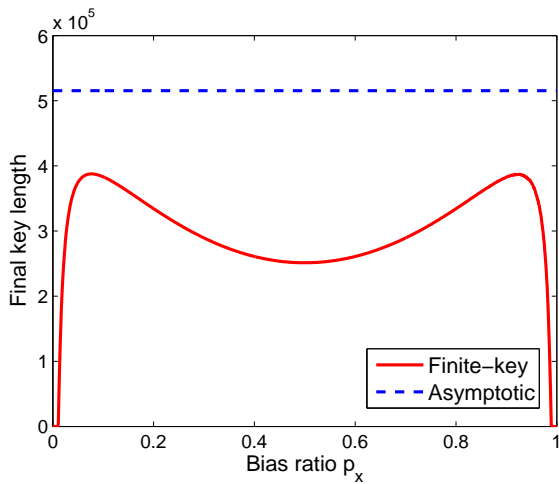}
\caption{\label{fig:Kbias}
Effect of the bias ratio on the final key length;
parameters used:
$e_{bx}=e_{bz}=4\%$, target failure probability $\varepsilon=10^{-7}$, the raw key length is $10^6$, and the error correction efficiency is 100\%.
}
\end{figure}
\begin{figure}
\includegraphics[width=\columnwidth]{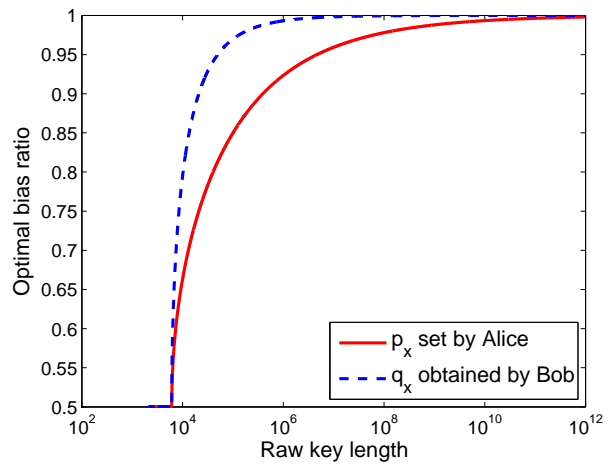}
\caption{\label{fig:OptbiasN}
Plot of the optimal bias ratio vs. the raw key length;
parameters used:
$e_{bx}=e_{bz}=4\%$, target failure probability $\varepsilon=10^{-7}$, and the error correction efficiency is 100\%.
}
\end{figure}
Figure~\ref{fig:Kbias} illustrates the effect of the bias ratio on the final key length.
It can be seen that when the optimal bias ratio is used, the final key length increases by over $50\%$ compared to the case when the bias ratio of $0.5$ is used.
Thus, the bias ratio has a big effect on the key generation performance.
Figure~\ref{fig:OptbiasN} shows the optimal bias ratio versus the raw key length.
The optimal bias ratio leads to the largest final key length.
It can be seen that as the raw key length approaches infinity, the optimal bias ratio tends to one.
This makes sense since in the asymptotic case, it is more efficient for Alice and Bob 
to use one basis with a high probability for key generation in order to avoid wasteful basis mismatch, and to use the other basis only 
for phase error estimation; this is the idea of the efficient BB84 protocol~\cite{EffBB84_05}.
The optimal bias ratio drops to $0.5$ as the raw key length approaches the minimal for positive key generation.

\section{Concluding remarks\label{sec-conclusion}}
In this paper, we propose a complete post-processing procedure for transforming measurement outcomes in a QKD experiment to a final secret key, which we quantify with a security parameter, namely the failure probability of the post-processing procedure.
This failure probability is directly connected to the composability security definition (cf. Lemma~\ref{lemma-relation-failure-prob-trace-distance}).
Our post-processing procedure contains many elements including authentication, the choice of the basis bias ratio, error correction and verification, phase error rate estimation, and privacy amplification.
Our procedure results from integrating all these elements with ideas from security proofs. 
Since the underlying security proofs  
\cite{LoChauQKD_99,ShorPreskill_00,Lo_decouple_2003,Koashi_Uncer_06} are secure against the most general attacks, our post-processing procedure also inherits this important property.
Based on our analysis, the main contribution to the finite-size effect comes from the inefficiency of phase error rate estimation, 
which is a consequence of 
the random sampling argument for inferring unobserved quantities from observed ones.
Further remarks and future directions are listed as follows:

\begin{enumerate}
\item
In the privacy amplification step, Alice and Bob need a common matrix to generate the final secure key. The current way to construct the matrix is by Alice sending a random bit string to Bob, which requires authenticated classical communication. An alternative way is by each of them generating a matrix with a pre-shared secret key. Of course, the amount of pre-shared secret key bits required must be small compared to the generated key length.  
Also, the failure probability is related to the amount of pre-shared bits consumed.
We leave this investigation for future research.  The main advantage of the second method is that no classical communication is needed for the privacy amplification part. In this case, the error verification step can be done either before or after the privacy amplification.

\item
In the security proof, 
the imperfection of $X$- and $Z$-basis measurements and efficiency mismatch are not considered.
It is interesting to consider the detector efficiency mismatch with the finite key analysis \cite{Mismatch_security_09}.

\item
As noted in Ref.~\cite{Practical_05}, the finite-key analysis for the decoy-state QKD is a hard problem. In the decoy-state QKD, the fluctuation comes from not only statistics but also hardware imperfections. The question of interest is where the main contribution of the fluctuation comes from and how to quantify these fluctuations. Since QKD systems with coherent states are most widely used in experiments, investigating the finite key effect in decoy-state QKD is an important step towards a QKD standard.

\item
In order to fairly compare our finite-key analysis to others, such as 
Scarani and Renner~\cite{Scarani:Finite:08} and Cai and Scarani~\cite{Cai_finite_2008}, 
one has to make sure the post-processing elements of different post-processing procedures carry similar capacities.
For example,
there are different ways to treat the basis bias ratio, authentication, and random sampling.
Therefore, a clear objective must be defined first before making a meaningful comparison.
We remark that 
comparing the performance of various post-processing procedures as a whole 
and
comparing only the underlying security proofs (which are just one element 
in a post-processing procedure)
are two different goals.
As we have shown in this paper, the main contribution to the finite-size effect comes from random sampling in the parameter estimation step.
Thus, 
it may be more interesting in practice to compare different random sampling arguments.

\item 
Our analysis treats the $X$-basis and $Z$-basis separately, especially when we estimate the phase error rates using the random sampling argument.
On the other hand, one may mix the measurement data of different bases before any analysis.
Doing so makes it easy to perform a similar finite-key analysis for other protocols such as the SARG04 protocol~\cite{SARG_04}.
In this case, we can use the Azuma's inequality~\cite{Azuma1967} in place of the random sampling argument to estimate the phase error rate.
This is discussed in more detail in Appendix~\ref{app-azuma}.

\item
In QKD experiments, error correction is often performed in blocks (say, 1
kbit) and privacy amplification is performed on all the blocks together. In
some error correction scheme, the failure probability for small blocks is
not negligible. That is, after the error correction, some blocks may still
have errors, discarding these blocks may have security implication and thus
care is required~\cite{Tittel2009}.
It is an interesting future topic to give a strict
security argument on this issue.

\item Although our analysis uses particular procedures for the steps (e.g., authentication, error correction), our analysis is generic in the sense that each specific procedure may be substituted by another with the same functionality.
The new secret-key cost and failure probability will then be used in the analysis of the generation rate and failure probability of the final key.

\end{enumerate}

\section{Acknowledgments}
We thank J.-C. Boileau, C.~Erven, N.~Godbout, M.~Hayashi, D.~W.~Leung, H.-K.~Lo, N.~L\"utkenhaus, M.~Koashi, X.~Mo, B.~Qi, R.~Renner, V.~Scarani, D.~Stebila, K.~Tamaki, W.~Tittel, Q.~Wang, Y.~Zhao and other participants in the workshop \emph{Quantum Works QKD Meeting (Waterloo, Canada)} and \emph{Finite Size Effects in QKD (Singapore)} for enlightening discussions. X.~Ma especially thanks H.~F.~Chau for hospitality and support during his visit at the University of Hong Kong. This work is supported by the NSERC Innovation Platform Quantum Works, the NSERC Discovery grant, the RGC grant No.~HKU 701007P of the HKSAR Government, and the Postdoctoral Fellowship program of NSERC.

\begin{appendix}

\section{Proof of Eq.~(\ref{eqn-fidelity-bound-general1})}\label{App:proof-fidelity-bound}

First, given that
$\langle 0_X^{\otimes m} | \rho_A | 0_X^{\otimes m}\rangle = \alpha$,
the purification 
$\ket{\Psi_{AE}}$ of $\rho_{A}$ is of the general form
\begin{align}
\ket{\Psi_{AE}} = \sqrt{\alpha} \ket{0_X^{\otimes m}}_A \ket{0}_E + \sqrt{1-\alpha} \ket{\Psi^\perp}_{AE}
\end{align}
where $\ket{\Psi^\perp}_{AE}$ has unit norm and 
$_A\langle 0_X^{\otimes m}\ket{\Psi^\perp}_{AE}=0$.
Thus, the fidelity in question is
\begin{align}
& F\left(\rho_{AE},\ket{0_X^{\otimes m}}_A \bra{0_X^{\otimes m}} \otimes \rho_E\right)
\nonumber \\
=&
\operatorname{Tr}\sqrt{
\ket{\Psi_{AE}}\bra{\Psi_{AE}}
\left[\ket{0_X^{\otimes m}}_A \bra{0_X^{\otimes m}} \otimes \rho_E \right]
\ket{\Psi_{AE}}\bra{\Psi_{AE}}
} \nonumber \\
=&
\operatorname{Tr}\sqrt{
\ket{\Psi_{AE}}
\left[\alpha  \bra{0} \rho_E \ket{0} \right]
\bra{\Psi_{AE}}
} \nonumber \\
=&
\sqrt{\alpha  \bra{0} \rho_E \ket{0} }
\label{eqn-fidelity-bound-proof-1}
\end{align}
where $\rho_{AE}=\ket{\Psi_{AE}}\bra{\Psi_{AE}}=\rho_{AE}^{1/2}$.
Since
$\rho_E=\operatorname{Tr}_A (\rho_{AE})$
and
\begin{align}
\bra{0} \rho_E \ket{0}
=
\sum_i | \bra{\Psi_{AE}} \left[ \ket{i}_A \ket{0}_E \right] |^2
\end{align}
where the summation is over all vectors of a basis in system $A$,
by considering a basis having $\ket{0_X^{\otimes m}}_A$ as its basis state, we get
$\bra{0} \rho_E \ket{0} \geq \alpha$.
Substituting this into Eq.~\eqref{eqn-fidelity-bound-proof-1}, we get Eq.~\eqref{eqn-fidelity-bound-general1}.

\section{Evaluation of hypergeometric function} \label{App:Hyper}
In this appendix, we will evaluate the hypergeometric function
\begin{equation} \label{App:Random:Hypergeo}
\begin{aligned}
P_{\theta} \le \textnormal{Pr}\{k|m,n,N\} = {{{m \choose k} {{N-m} \choose {n-k}}}\over {N \choose n}}.
\end{aligned}
\end{equation}
with $k=e_{bx}n_x$, $N=n_x+n_z$, $n=n_x$ and $m=e_{bx}(n_x+n_z)+\theta n_z$. Here, we relabel the function for simplicity. Strictly speaking, $\theta$ is a discrete variable with a minimum quantum of $1/n_z$, which will keep $m$ to be an integer. 

In the following discussion, we assume the integers $N>m>k\ge1$ and $N>n>k$. The only exception that could (though highly unlikely) happen in the realistic case is $k=0$. In this case, for a given $m$, $P_{\theta}(k=0)<P_{\theta}(k=1)$. Now that we only care about the upper bound of the probability, we can always safely replace $k=0$ with $k=1$ in the calculation.

We simplify the hypergeometric function by the Stirling formula \cite{Robbins:Stirling:55}
\begin{equation} \label{Random:Stirling}
\begin{aligned}
n! = \sqrt{2 \pi n} \left(\frac{n}{e}\right)^{n}e^{\lambda_n},
\end{aligned}
\end{equation}
where
\begin{equation} \label{Random:StFactor}
\begin{aligned}
\frac{1}{12n+1} < \lambda_n < \frac{1}{12n}.
\end{aligned}
\end{equation}
Then, Eq.~\eqref{App:Random:Hypergeo} can be expressed as
\begin{equation} \label{Random:HyperStir}
\begin{aligned}
P_{\theta}
&\le {{{n \choose k} {{N-n} \choose {m-k}}}\over {N \choose m}} \\
&=  { n!(N-n)!(N-m)!m! \over k!(n-k)!(m-k)!(N-n-m+k)!N! } \\
&= \frac1{\sqrt{2 \pi}} \frac{\sqrt{n}\sqrt{N-n}\sqrt{N-m}\sqrt{m}} {\sqrt{k}\sqrt{n-k}\sqrt{m-k}\sqrt{N-n-m+k}\sqrt{N}} \\
&~~~\cdot \frac{n^n(N-n)^{N-n}(N-m)^{N-m}m^{m}} {k^{k}(n-k)^{n-k}(m-k)^{m-k}} \\
&~~~\cdot \frac{1}{(N-n-m+k)^{N-n-m+k}N^{N}} \\
&~~~\cdot \exp(\lambda_n+\lambda_{N-n}+\lambda_{N-m}+\lambda_{m} -\lambda_k- \\
&~~~~~~~~~~~
\lambda_{n-k}-\lambda_{m-k}-\lambda_{N-n-m+k}-\lambda_{N}) .
\end{aligned}
\end{equation}

First, we can prove that
\begin{equation} \label{Random:HyperLamdas}
\begin{aligned}
\lambda_n+&\lambda_{N-n}+\lambda_{N-m}+\lambda_{m} -\lambda_k-\\
&\lambda_{n-k}-\lambda_{m-k}-\lambda_{N-n-m+k}-\lambda_{N} &< 0 
\end{aligned}
\end{equation}
with the facts of $m>k\ge 1$, $n-k>1$ and Eq.~\eqref{Random:StFactor}. Remark: though the left-hand side of Eq.~\eqref{Random:HyperLamdas} is negative, it is close to 0 in the order of $O(1/12k)$.

Second, we know that ${1}/{\sqrt{x(1-x)}}$ is a decreasing function for $0<x<1/2$. Then we can easily see that
\begin{equation} \label{Random:HyperSqrts}
\begin{aligned}
&\frac{\sqrt{n}\sqrt{N-n}\sqrt{N-m}\sqrt{m}} {\sqrt{k}\sqrt{n-k}\sqrt{m-k}\sqrt{N-n-m+k}\sqrt{N}} \\
\le & \frac{\sqrt{N}}{\sqrt{n(N-n)}}\frac{1}{\sqrt{e_{bx}(1-e_{bx})}} \\
=& \frac1{\sqrt{N}}\frac{1}{\sqrt{q_x(1-q_x)e_{bx}(1-e_{bx})}} 
\end{aligned}
\end{equation}
with the facts of $e_{bx}=k/n$ and $e_{pz}=(m-k)/(N-n)\ge e_{bx}$. Remark: when $e_{pz}=e_{bx}$, the equality holds. From this point of view, the bound is tight.

Third, the remaining term of the failure probability can be expressed by
\begin{equation} \label{Random:HyperExp}
\begin{aligned}
&\frac{n^n(N-n)^{N-n}(N-m)^{N-m}m^{m}} {k^{k}(n-k)^{n-k}(m-k)^{m-k}(N-n-m+k)^{N-n-m+k}N^{N}} \\
&= 2^{nH(\frac{k}{n})+(N-n)H(\frac{m-k}{N-n})-NH(\frac{m}{N}) } \\
&\equiv 2^{ -N\xi_x(\theta) } \\
\end{aligned}
\end{equation}
where we use the definition of the binary entropy function $H(x)$. The exponent coefficient is given by
\begin{equation} \label{App:Random:XiPhase}
\begin{aligned}
\xi_x(\theta) \equiv & H(e_{bx}+\theta-q_x\theta)-q_xH(e_{bx})\\
& -(1-q_x)H(e_{bx}+\theta) 
\end{aligned}
\end{equation}
with $q_x=n/N$ and $({m-k})/({N-n})=e_{bx}+\theta$. Due to the concavity of $H(x)$, $\xi_x(\theta)$ is negative for $\theta>0$ and $0<q_x<1$.

Therefore, by combining Eqs.~\eqref{Random:HyperStir}, \eqref{Random:HyperLamdas}, \eqref{Random:HyperSqrts} and \eqref{Random:HyperExp}, the failure probability of Eq.~\eqref{Random:Pbound} is given by
\begin{equation} \label{App:Random:FailApp}
\begin{aligned}
P_{\theta} &<
\frac1{\sqrt{N}}\frac{1}{\sqrt{q_x(1-q_x)e_{bx}(1-e_{bx})}} 2^{ -N\xi_x(\theta) } \\
\end{aligned}
\end{equation}
where $\xi_x(\theta)$ is given by 
Eq.~\eqref{App:Random:XiPhase}.
Note that $\xi_x(\theta)$ is independent of key size $N$ given the error rates and bias ratio. Now we can see that the failure probability decreases (actually, slightly faster than) exponentially with $N$.

\section{Proof of Eq.~(\ref{eqn-number-of-phase-errors-binomial})\label{app-claim-binomial-sum}}

We prove Eq.~\eqref{eqn-number-of-phase-errors-binomial} by the following claim.
\begin{claim}
\begin{equation} 
\begin{aligned}
\sum_{k=0}^{m-1}{{n} \choose {k}} < {{n} \choose {m}}
\end{aligned}
\end{equation}
\end{claim}
when $m\le n/3$.
\begin{proof}
First notice that
\begin{equation} 
\begin{aligned}
\frac{{{n}\choose{k-1}}}{{{n}\choose{k}}} &= \frac{k}{n-k+1} < \frac12
\end{aligned}
\end{equation}
is true for all $k\le n/3$. Thus,
\begin{equation} 
\begin{aligned}
\sum_{k=0}^{m-1}{{n} \choose {k}} &\le \sum_{k=0}^{m-1}2^{k-m}{{n} \choose {m}} \\
&= {{n} \choose {m}} \sum_{k=0}^{m-1}2^{k-m} \\
&< {{n} \choose {m}} \\
\end{aligned}
\end{equation}
is true for $m\le n/3$.
\end{proof}

\section{Estimation of phase error rate for mixed-basis analysis\label{app-azuma}}

The analysis in the main part of the paper treats each of the two bases separately when estimating the phase error rates for them.
This is possible in BB84, since
the phase errors in one basis are the bit errors in the other basis.
And in this case,
a random sampling argument suffices to establish some confidence on the unmeasured phase error rate in one basis from the measured bit error rate in the other basis.
On the other hand, one may mix all the measurement data in the different bases together before applying any further analysis.
This can be done in BB84.
For other protocols, this mixing actually leads to a simpler analysis and thus is favorable.
Here, we describe how to estimate the phase error rate for the mixed-basis case.
When the measurements are mixed, protocols can usually be characterized with a 
relation between the bit and phase error probabilities
$p_{p}=\alpha p_{b}$, where $\alpha\geq 1$ in general (e.g., $\alpha=3/2$ for SARG04~\cite{TamakiLo_06,FTL_06} and $\alpha=5/4$ for a three-state protocol~\cite{Boileau2005}).
(Note that here the error probabilities are the combined values of all bases and thus do not carry a basis designation.)
Given such a relation in probabilities, we want to establish a similar relation for the error rates and compute the confidence for it.
A useful tool to do this is the Azuma's inequality~\cite{Azuma1967} (see also Refs.~\cite{Boileau2005,TamakiLo_06,FTL_06,Fung2006b} for the application of it to security proofs), which relates the sum of conditional probabilities to the total number of a particular outcome in many trials.
To start, we relate the probability and the rate for the bit error and the phase error separately using the Azuma's inequality as follows:
\begin{align}
\label{eqn-Azuma-bit}
\text{Pr} \{ |p_{b} - e_{b}| \geq \varepsilon_\text{Az} \} \leq 2 \exp (\frac{-n \varepsilon_\text{Az}^2}{2})\\
\label{eqn-Azuma-phase}
\text{Pr} \{ |p_{p} - e_{p}| \geq \varepsilon_\text{Az} \} \leq 2 \exp (\frac{-n \varepsilon_\text{Az}^2}{2})
\end{align}
where $p_{b,p}$ and $e_{b,p}$ designate the error probabilities and the error rates respectively,  $\varepsilon_\text{Az}$ represents a failure probability, and $n$ is the number of measurements made.
Because $p_{p}=\alpha p_{b}$, we multiple these two inequalities to get the relation between the bit and phase error rates:
\begin{align}
\text{Pr} \{ |e_{p} - \alpha e_{b}| \geq (1+\alpha)\varepsilon_\text{Az} \} \leq 4 \exp (-n \varepsilon_\text{Az}^2) .
\end{align}
For BB84, $\alpha=1$ and this bound is worse than the random sampling result (cf. Eq.~\ref{Random:FailTheta}) in typical situations.

\end{appendix}

\bibliographystyle{apsrev}

\bibliography{Bibli30,../ref}

\begin{thebibliography}{78}
\bibitem[{Tit({\natexlab{b}})}]{Tittel2009}
\bibinfo{note}{This issue was raised by Wolfgang Tittel}.
\end{thebibliography}

\end{document}